\documentclass[preprint,pteplogo]{ptephy_v2}

\preprintnumber{XXXX-XXXX} 
\usepackage{hyperref}
\usepackage{enumerate}
\usepackage{comment}
\usepackage{bm}
\usepackage{physics}
\usepackage{subfig}
\usepackage{amsmath}
\usepackage{here}

\begin{document}

\title{Structure and Formation of the Deeply Bound $\bar{p}$ atoms}


\author{N. Miyazaki}
\affil{Department of Physics, Kyoto Sangyo University, Kyoto 603-8555, Japan \email{i2585147@cc.kyoto-su.ac.jp}}

\author[1]{J. Yamagata-Sekihara}

\author{S. Hirenzaki}
\affil{Department of Physics, Nara Women's University, Nara 630-8506, Japan}




\begin{abstract}%
We study  theoretically the structure and formation of the deeply bound $\bar{p}$ atoms.
We find that the widths of the atomic states are narrower than the level spacing even for deeply bound states so that the well-isolated deeply bound $\bar{p}$ atoms are expected to exist.
We also find the $\bar{p}$-nuclear states with huge widths.
For the observation of the deep $ {\bar p}$-atomic states, we investigate theoretically the $(\bar{p}, p)$ reactions for $^{12}$C, $^{16}$O, and $^{31}$P target nuclei.
We find that the momentum transfer of the $( {\bar p},p)$ reaction is small and the formation of the $ {\bar p}$-atomic states can be observed as the discrete peak structures in the $( {\bar p},p)$ spectrum.
We conclude that the $(\bar{p}, p)$ reactions are very much suited for the $\bar{p}$ atom formation and the spectra of the reaction are expected to provide new valuable information on the $ {\bar p}$ atoms and $ {\bar p}$-nucleus interaction.
\end{abstract}

\subjectindex{xxxx, xxx}

\maketitle

\section{Introduction}
Exotic atoms, which are the bound states of a negatively charged hadron and a nucleus, have been studied for decades to study the various aspects of the exotic hadron many body systems.
So far, considerable information on hadron-nucleus interactions and structures of the exotic atoms has been accumulated \cite{Batty:1997zp,Friedman:2007zza}, and the information is of great interest because of its essential importance for understanding the nature of hadrons at finite density \cite{Hayano:2008vn,Yamazaki:2012zza,Metag:2017yuh}.
One of the important advantages of the studies of the exotic atoms is that we can determine the strength of each potential term with various properties by the precise spectroscopic studies of the bound states with different quantum numbers, based on the different sensitivities of the potential terms to the observables.
The exclusive information is very important to deduce the microscopic meanings of the terms.
For example, the precise determination of one potential parameter $(b_{1})$ of the pion-nucleus optical potential enables us to determine the accurate value of the $\ev*{{\bar q}q}$ condensation at finite density by the study of deeply bound pionic atoms \cite{Suzuki:2002ae,piAF:2022gvw}.

By the precise studies of the ${\bar p}$-nucleus bound systems, we can expect to determine the ${\bar p}$-nucleus interaction precisely for various types of potential terms such as $s$-wave, $p$-wave, isoscalar, isovector, and so on.
The strength of these potential terms may provide the information on the ${\bar p}$-$p$ and ${\bar p}$-$n$ interactions at finite density.
We can also expect to deduce the information on the nuclear surface from the $ {\bar p}$ atom data \cite{Yoshimura:2024oza}.
In addition, it may also be possible to produce even more exotic hadron systems from the decay of ${\bar p}$ atom by the ${\bar p}N \to$ hadrons processes \cite{Amsler:1997up,Ohnishi:2014xla} similar to the hyper nucleus formation from the kaonic atom decay \cite{Matsuyama:1988ep} .
We think the systematic studies of the ${\bar p}$ atoms provide the important information on the interactions between the hadron and antihadron and the hints to get deeper insights for the origin of the matter-antimatter asymmetry in our universe.

In this article, we study the structure and formation of the antiprotonic atoms (${\bar p}$ atoms) including the deeply bound states which cannot be observed by the X-ray spectroscopy.
Because of the strong nuclear absorption of $ {\bar p}$ in the deexcitation processes, the last orbital of $ {\bar p}$ atom, which is the deepest state observed by the X-ray spectroscopy, is $3d$ for C and O, and is $4f$ for Si, for example \cite{Batty:1995am}.
The recent studies of the ${\bar p}$ atoms were reported in Refs.~\cite{Batty:1997zp,Friedman:2007zza,Friedman:1999pu} where the several phenomenological ${\bar p}$-nucleus interactions are obtained by the experimental data.
We use one of the potential in Ref.~\cite{Batty:1997zp} and study the structure of the $ {\bar p}$-nucleus bound systems.
We also study the $({\bar p},p)$ reactions for the formation of ${\bar p}$ atoms theoretically to extend the spectroscopic study to new frontier of deeply bound ${\bar p}$ atoms.
The possibility of the existence of deeply bound $ {\bar p}$ atomic states has already been discussed in early theoretical studies.
In particular, Gibbs and Kaufmann \cite{Gibbs:1984rr} investigated deeply bound $1s$ $ {\bar p}$ atomic states in medium-mass nuclei, while Baltz et al. \cite{Baltz:1985zz} studied $ {\bar p}$-nucleus interactions in atomic systems.
These studies mainly focused on estimating the formation strengths or cross sections of the bound states, as well as their binding energies and widths.
Some simple estimations of the $ {\bar p}$ atom formation are also given in Ref.~\cite{Friedman:1999pu}.
As described here in the last paragraph in sect.~\ref{subsec:Neff}, the forward $({\bar p},p)$ reactions are almost recoilless for any initial energies and are considered to be very suited for the formation of the so-called substitutional $ {\bar p}$-atomic state with a proton hole.

\section{Theoretical formalism}
\subsection{Structure of the $\bar{p}$ atom}
\begin{figure}[!ht]
\centering
\includegraphics[width=0.6\columnwidth]{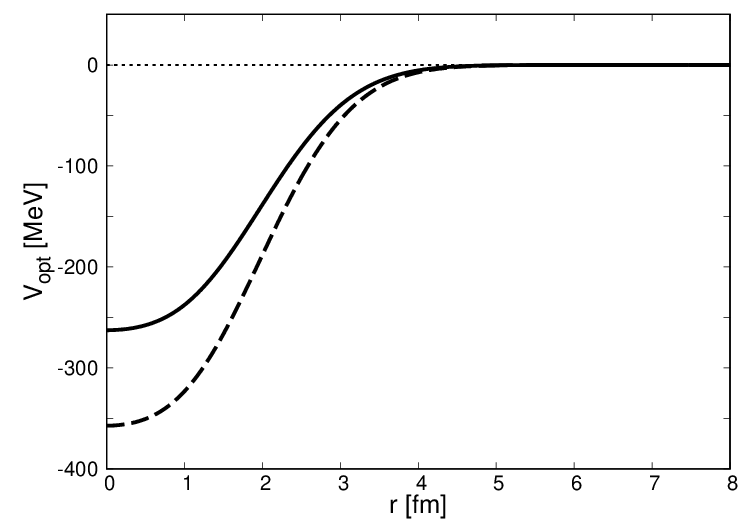}
\caption{The antiproton-$^{11}$B optical potential as functions of the radial coordinate $r$.
The potential is determined in Ref.~\cite{Batty:1997zp} as shown in Eqs.~\eqref{eq:Optical} and \eqref{eq:b0} in the text.
The solid and dashed lines show the real and imaginary parts of the potential, respectively.}
\label{fig:Optical_11B}
\end{figure}

The $\bar{p}$-nucleus optical potential $V_{\rm{opt}}$ used for the calculations of bound systems and formation spectra of the $ {\bar p}$-nuclear systems is given by
\begin{equation}\label{eq:Optical}
	2\mu V_{\text{opt}}(r) = -4\pi\eta b_0 \rho(r),
\end{equation}
where $\mu$ is the reduced mass of the antiproton and the nucleus, $\rho(r)$ the nuclear density distribution, and $b_0$ the complex effective scattering length which determine the potential strength.
$\eta$ is defined as $\eta=1+m_{\bar{p}}/M_N$ with antiproton ($m_{\bar p}$) and nucleon ($M_N$) masses.
The value of $b_0$ is obtained in a phenomenological fit to the experimental data in Ref.~\cite{Batty:1997zp}.
We use one of the results reported in Ref.~\cite{Batty:1997zp} in this article, as
\begin{equation}\label{eq:b0}
b_0=(2.5\pm0.3)+i(3.4\pm0.3) ~{\rm fm}.
\end{equation}
As for the nuclear density distribution $\rho$, we adopt the functional form of the hormonic-oscillator and three-parameter Fermi distributions as shown in Table~\ref{table:ADATA} with their parameter values which are determined in Refs.~\cite{DeJager:1974liz,DeVries:1987atn} for the nuclear charge distributions.
\begin{table}[b]
\caption{The parameters of the harmonic-oscillator (HO) $\rho(r)=\rho_0(1+\alpha(r/a)^2)\exp(-(r/a)^2)$ and three-parameter Fermi (3pF) $\rho(r)=\rho_0(1+wr^2/c^2)/(1+\exp((r-c)/z))$ nuclear density distributions \cite{DeJager:1974liz,DeVries:1987atn} for nuclei considered in this article.}
\label{table:ADATA}
\centering
\begin{tabular}{c|c|cccc}
\hline
Nucleus & Density & $c$ [fm] or $a$ [fm] & $z$ [fm] or $\alpha$ & $w$ & Ref.\\
\hline
$^{11}$B & HO & 1.69 & 0.811 & & \cite{DeJager:1974liz}\\
$^{12}$C & HO & 1.687 & 1.067 & & \cite{DeJager:1974liz}\\
$^{15}$N & 3pF & 2.334 & 0.498 & 0.139 & \cite{DeJager:1974liz}\\
$^{16}$O & 3pF & 2.608 & 0.513 & -0.051 & \cite{DeJager:1974liz}\\
$^{30}$Si & 3pF & 3.252 & 0.553 & -0.078 & \cite{DeVries:1987atn}\\
$^{31}$P & 3pF & 3.369 & 0.582 & -0.173 & \cite{DeVries:1987atn}\\
\hline
\end{tabular}
\end{table}
As an example, we show the optical potential of $ {\bar p}$ for $^{11}$B nucleus in Fig.~\ref{fig:Optical_11B}.
The optical potential has the strong attractive real part and the even stronger absorptive imaginary part due to the antiproton annihilation processes with nucleon.

\begin{figure}[h]
\centering
\includegraphics[width=0.6\columnwidth]{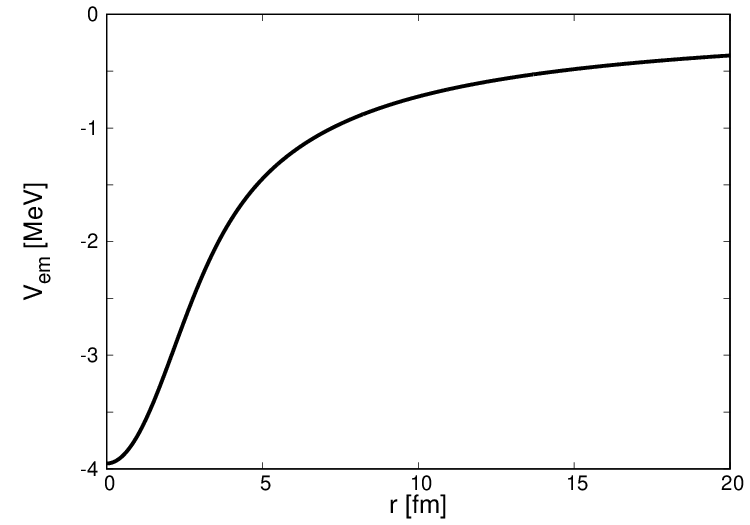}
\caption{The antiproton-$^{11}$B electromagnetic potential $V_{\rm em}(=V_{\rm FS}+V_{\rm VP})$ described in Eqs.~\eqref{eq:Coulomb_FS} and \eqref{eq:Coulomb_VP} as the function of the radial coordinate $r$.}
\label{fig:Coulomb_11B}
\end{figure}

It is also necessary to consider the realistic electromagnetic interaction between ${\bar p}$ and nucleus.
In this article, we include the leading-order vacuum polarization effects to the Coulomb potential, and take into account the finite charge density distribution of nucleus.
The finite size Coulomb potential $V_{\rm FS}(r)$ is given by
\begin{equation}\label{eq:Coulomb_FS}
	V_{\text{FS}}(r) = -\alpha \int \frac{\rho_{\rm ch}(r')}{\abs{\bm{r}-\bm{r'}}} \dd^3r',
\end{equation}
where $\alpha$ is the fine structure constant, and $\rho_{\rm ch}$ the charge density distribution which is normalized to the proton number of nucleus.
The charge distribution $\rho_{\rm ch}$ is taken to be the same form as the nuclear density $\rho$ and is obtained by the same parameters and functional forms as listed in Table~\ref{table:ADATA}.
The leading-order vacuum polarization effects with the finite charge distribution $V_{\rm VP}(r)$ can be calculated from Eq.~(7.24) in Ref.~\cite{Itzykson:1980rh}, and defined as
\begin{equation}\label{eq:Coulomb_VP}
	V_{\text{VP}}(r) = -\frac{2\alpha^2}{3\pi} \int \dd^3r' \frac{\rho_{\rm ch}(r')}{\abs{\bm{r}-\bm{r'}}} 
							\int_1^\infty \dd u e^{-2m_e\abs{\bm{r}-\bm{r'}}u} \qty(1+\frac{1}{2u^2})\frac{\sqrt{u^2-1}}{u^2},
\end{equation}
where $m_e$ is the electron mass.
The total electromagnetic interaction $V_{\rm em}(r)$ is given by the sum of $V_{\rm FS}$ and $V_{\rm VP}$ of Eqs.~\eqref{eq:Coulomb_FS} and \eqref{eq:Coulomb_VP} as $V_{\rm em} = V_{\rm FS} + V_{\rm VP}$ \cite{Ikeno:2015ioa}.
We show in Fig.~\ref{fig:Coulomb_11B} the electromagnetic potential $V_{\rm em}(r)$ for antiproton-$^{11}$B system as an example.

In the current exploratory level, the effects of the spin of the antiproton in its atomic state are not significant and are often dismissed from considerations.
In fact, the potential parameter in Eq.~\eqref{eq:b0} is determined by comparing the calculated values using the Klein-Gordon equation with the experimental data \cite{Batty:1997zp}.
Thus, we follow here the prescription in Ref.~\cite{Batty:1997zp} and employ the Klein-Gordon equation
\begin{equation}\label{eq:KGeq}
	[-\laplacian+\mu^2+2\mu V_{\text{opt}}(r)]\phi(\bm{r})=[E-V_{\text{em}}(r)]^2\phi(\bm{r})
\end{equation}
as the equation of motion of antiprotonic atoms.
The complex wave functions $\phi({\bm r})$ and the complex eigenenergies $E$ of antiprotonic atoms can be obtained by solving Eq.~\eqref{eq:KGeq} with the boundary conditions of the bound states.
The binding energies $B_{\bar p}$ and the widths $\Gamma_{\bar p}$ of the atomic states are defined as $B_{\bar p}=\mu-{\Re}~E$ and $\Gamma_{\bar p}=-2\times{\Im}~E$.

\subsection{$(\bar{p},p)$ reaction for the $\bar{p}$ atom formation} \label{subsec:Neff}
In this subsection, we explain the $(\bar{p},p)$ reactions for the formation of the antiproton-nucleus bound systems based on the formalism of the effective number approach.
This approach is known to be one of the most convenient way to get numerical results for the formation cross sections of various bound states \cite{Yamazaki:2012zza,Yamagata:2005ic}.
In this approach, the emitted proton energy spectra at the fixed angle can be written as
\begin{equation}\label{eq:ENA1}
	\frac{\dd^2\sigma}{\dd E_{p}\dd\Omega_{p}} = \qty(\frac{\dd\sigma}{\dd\Omega})^{\text{ele}}_{\bar{p}p\to p\bar{p}}
	\sum N_{\text{eff}}~ \frac{\Gamma_{\bar p}}{2\pi} \frac{1}{\Delta E^2 + \Gamma_{\bar p}^2/4},
\end{equation}
where the sum is taken over all $({\bar p}$ state$)\otimes({\rm proton}$-${\rm hole})$ configurations for the final states with one $ {\bar p}$-particle and one proton-hole.
The differential cross section $(\dd\sigma/\dd\Omega)^{\rm ele}_{\bar{p}p\to p\bar{p}}$ of the elementary process ${\bar p}+p\to p+{\bar p}$ in the laboratory frame is obtained from the data reported in Ref.~\cite{Yoh:1969td}.
The energies of the resonance peaks for all configurations of the $ {\bar p}$-particle and $p$-hole in the final state are determined by the condition of $\Delta E=0$ in Eq.~\eqref{eq:ENA1}.
$\Delta E$ is defined as
\begin{equation}
	\Delta E=T_p-(T_{\bar p}-S_p(j_p^{-1})+B_{\bar p}),
\end{equation}
where $T_{\bar p}$ is the incident antiproton kinetic energy, $T_p$ the emitted proton kinetic energies, and $B_{\bar p}$ the antiproton binding energies in the final states.
The proton separation energies $S_p(j_p^{-1})$ from each single particle level denoted by $j_p$ are obtained from the experimental data and compiled in Table~\ref{table:Sp}.
The widths of the peaks are determined by the widths of the $ {\bar p}$-bound states $\Gamma_{\bar p}$.
The effects of the widths of the $p$-hole state $\Gamma_{p}$ are also taken into account later by folding the spectrum with the appropriate distribution function.
\begin{table}[b]
\caption{One proton separation energies $S_p$ and widths $\Gamma_p$ of the hole states of $^{12}$C and $^{16}$O deduced from the data in Ref.~\cite{Belostotskii:1985}  for $^{12}$C, and those in Refs.~\cite{Yosoi:2004whb} and \cite{Ajzenberg-Selove:1991rsl} for $^{16}$O. The separation energies corresponding to the ground states of the daughter nuclei are taken from Ref.~\cite{Firestone}. The widths $\Gamma_p$ indicate FWHM of the Gaussian distribution for $^{12}$C and of the Lorentz distribution for $^{16}$O. The widths of the ground states of the daughter nuclei are fixed to zero, because of their stabilities.}
\label{table:Sp}
\centering
\begin{tabular}{|c|cc|cc|}
\hline
single particle & \multicolumn{2}{c|}{$^{12}$C} & \multicolumn{2}{c|}{$^{16}$O}\\
states [MeV] & $S_p$ & $\Gamma_p$ & $S_p$ & $\Gamma_p$\\
\hline
$1p_{1/2}$ & & & 12.1 & 0\\
$1p_{3/2}$ & 16.0 & 0 & 18.4 & $3.1\times10^{-6}$\\
$1s_{1/2}$ & 33.9 & 12.1 & 41.1 & 19.0\\
\hline
\end{tabular}
\end{table}

The effective number $N_{\rm eff}$ is defined as
\begin{align}\label{eq:Neff}
	N_{\text{eff}} &= \sum_{JMm_s} \abs{\int \chi^*_{p}(\bm{r})\xi^\dagger_{\frac{1}{2}m_s}[\phi^*_{l_{\bar{p}}}(\bm{r})\otimes\psi_{j_p}(\bm{r})]_{JM}\chi_{\bar{p}}(\bm{r})\dd^3r}^2,
\end{align}
where the proton and the antiproton wave functions in bound states are denoted by $\psi_{j_p}$ and $\phi^*_{l_{\bar p}}$.
We adopt the harmonic oscillator wave function for $\psi_{j_p}$.
As mentioned above, the spin of ${\bar p}$ is neglected in this article and ${\bar p}$ is treated here as a boson again.
The spin wave function of proton in the final state is denoted by $\xi_{\frac{1}{2}m_s}$.
$\chi_{\bar p}$ and $\chi_{p}$ are the initial and finial distorted waves of the projectile (${\bar p}$) and ejectile ($p$), respectively.
We use the eikonal approximation and replace $\chi_{\bar p}$ and $\chi_{p}$ as,
\begin{equation}
	\chi_{p}^*(\bm{r})\chi_{\bar p}(\bm{r})=\exp(i{\bm q}\cdot{\bm r})D(z,{\bm b}),
\end{equation}
where $\bm{q}$ is the momentum transfer between the projectile and ejectile and $D(z,{\bm b})$ the distortion factor  defined as
\begin{equation}
	D(z,\bm{b}) = \exp[-\frac{1}{2}\sigma_{\bar{p}N} \int_{-\infty}^z\dd z' \rho_A(z',\bm{b})-\frac{1}{2}\sigma_{pN} \int_z^{\infty}\dd z'\rho_{A-1}(z',\bm{b})]
\end{equation}
with the antiproton-nucleon ($\sigma_{{\bar p}N}$) and proton-nucleon ($\sigma_{pN}$) total cross sections.
$\rho_A(z',\bm{b})$ and $\rho_{A-1}(z',\bm{b})$ are the density distributions of the target and daughter nuclei at the beam direction coordinate $z$ and the impact parameter ${\bm b}$.
$\sigma_{ {\bar p}N}$ and $\sigma_{pN}$ are determined by the experimental data, and the value of $\sigma_{{\bar p}N}$ is fixed to be $\sigma_{{\bar p}N} = 117.42$ mb based on the experimental data of ${\bar p}$ and proton cross section $\sigma_{{\bar p}p}=117.42$ mb at $p_{\rm lab}=1$ GeV/c reported in Ref.~\cite{Workman:2022ynf} with assuming the same value for the ${\bar p}$ and neutron cross section $\sigma_{ {\bar p}n}$.
$\sigma_{pN}$ is also determined by the data in Ref.~\cite{Workman:2022ynf} as the weighted average of the proton-proton and the proton-neutron cross sections for each daughter nucleus.

Finally, we would like to mention here the special features of the (${\bar p}$,$p$) reaction for the formation of the ${\bar p}$-bound atomic states.
Thanks to two facts, (i) the masses of the projectile (${\bar p}$) and ejectile ($p$) are exactly same and (ii) the reaction Q-value is very small in the present reaction because of no new hadron production, the momentum transfer ${\bm q}$ is close to 0 for all incident energies at forward angles.
Namely, we can select the best incident energy to optimize the experimental conditions with keeping the recoilless condition unlike in the case of the $\pi$ atom formation \cite{Yamazaki:2012zza,Suzuki:2002ae,piAF:2022gvw}.
Thus, in principle, the (${\bar p}$,$p$) reaction has the advantage for the formation of the so-called substitutional states of $ {\bar p}$ atoms.
In addition, the angular dependence of the spectrum is expected to be suited for the observation of other final state configurations by changing the momentum transfer from almost zero to finite values.

\section{Numerical results}
\begin{figure}[htbp]
\centering
\includegraphics[width=0.7\columnwidth]{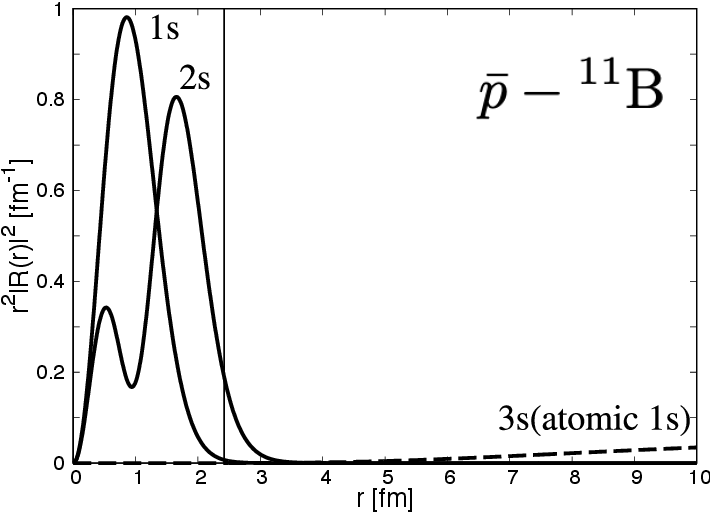}
\caption{Density distributions of ${\bar p}$ in the $s$-states of ${\bar p}-^{11}$B bound system. 
The solid lines represent the densities of 1st and 2nd $s-$states ($1s$ and $2s$) which are considered as the nuclear states.
The dashed line shows the density of the 3rd $s-$state ($3s$), which is interpreted as the atomic 1$s$ state.
The vertical solid line indicates the root-mean-square radius $\ev*{r^{2}}^{1/2}=2.42$ fm of $^{11}$B.}
\label{fig:11B_pbar_WF_N1sN2s1s}
\end{figure}

We first show the calculated antiproton-nucleus binding energies and widths for both $ {\bar p}-$nuclear and $-$atomic states in $^{11}$B, $^{15}$N, and $^{30}$Si in Table~\ref{table:BE-W}.
These nuclei are selected as the final state in the (${\bar p}$,$p$) reactions for the formation of the ${\bar p}$ bound states on the $^{12}$C, $^{16}$O, and $^{31}$P targets.
In all cases, plenty of bound states are found and some of deeply bound states are compiled in Table~\ref{table:BE-W}.
We first mention that the bound states can be classified clearly into $ {\bar p}-$nuclear states and $ {\bar p}-$atomic states by noting the features of the density distributions of antiproton as shown in Fig.~\ref{fig:11B_pbar_WF_N1sN2s1s} as a typical case.
As you can see in the figure, the density distributions of the nuclear states are confined almost inside the nucleus, while those of the atomic states have much larger spatial dimension and extend far beyond the nuclear radius.
These nuclear states and atomic states appear as a series of bound state solutions of the equation of motion with the same angular momentum, thus can not be distinguished by quantum numbers.
We classify these states by the density distributions, and express in Table~\ref{table:BE-W} the first and second $s$-states as nuclear $1s$ and $2s$ states for $^{11}$B, and the third and later states as atomic $1s$, $2s$, $\cdots$ states.
The coexistence of the nuclear and atomic bound states are known to exist for the systems with the attractive strong and electromagnetic interactions such as kaonic nuclei and kaonic atoms \cite{Yamagata:2005ic}.

\begin{table}[b]
\caption{Binding energies $B_{\bar p}$ and widths $\Gamma_{\bar p}$ of antiproton$-^{11}$B, $-^{15}$N and $-^{30}$Si bound systems in units of MeV for nuclear states and of keV for atomic states calculated with the phenomenological optical potential in Ref.~\cite{Batty:1997zp} shown in Eqs.~\eqref{eq:Optical} and \eqref{eq:b0} in the text.}
\label{table:BE-W}
\centering
\begin{tabular}{c|cc|cc|cc}
\hline\hline
Nuclear State & \multicolumn{2}{c|}{$^{11}$B} & \multicolumn{2}{c|}{$^{15}$N} & \multicolumn{2}{c}{$^{30}$Si}\\
(MeV) & $B_{\bar p}$ & $\Gamma_{\bar p}$ & $B_{\bar p}$ & $\Gamma_{\bar p}$ & $B_{\bar p}$ & $\Gamma_{\bar p}$ \\
\hline
1s & 109.2 & 722.8 & 121.9 & 667.3 & 142.0 & 712.3\\
2s & 12.4 & 504.7 & 36.5 & 517.0 & 80.9 & 602.5\\
3s & - & - & - & - & 8.2 & 485.0\\
2p & 62.1 & 615.6 & 84.6 & 599.7 & 115.4 & 661.2\\
3p & - & - & - & - & 46.7 & 546.5\\
3d & 12.5 & 513.4 & 43.0 & 530.4 & 85.7 & 609.2\\
4d & - & - & - & - & 11.4 & 491.0\\
4f & - & - & - & - & 53.5 & 557.0\\
\hline
Atomic State & & & & & & \\
(keV) & & & & & & \\
\hline
1s & 268.9 & 57.9 & 430.9 & 125.5 & 1100.8 & 323.1\\
2s & 94.3 & 12.4 & 164.6 & 30.6 & 493.1 & 101.1\\
3s & 47.7 & 4.5 & 86.5 & 11.8 & 280.9 & 44.2\\
4s & 28.7 & 2.1 & 53.3 & 5.7 & 181.5 & 23.1\\
2p & 139.3 & 5.3 & 264.6 & 25.7 & 853.8 & 160.4\\
3p & 62.3 & 1.8 & 119.8 & 8.7 & 412.7 & 57.7\\
4p & 35.2 & $8.0\times10^{-1}$ & 68.3 & 3.9 & 244.8 & 27.1\\
3d & 63.8 & $1.5\times10^{-2}$ & 127.9 & $2.9\times10^{-1}$ & 516.8 & 21.0\\
4d & 35.8 & $8.9\times10^{-3}$ & 71.9 & $1.7\times10^{-1}$ & 290.3 & 11.3\\
4f & 35.8 & $5.2\times10^{-6}$ & 71.9 & $3.4\times10^{-4}$ & 297.3 & $2.9\times10^{-1}$\\
\hline
\end{tabular}
\end{table}

As for the level structure of the $s-$states of ${\bar p}-^{11}$B, for example, we found that the level spacing between the nuclear $1s$ and $2s$ states is large as 96.8 MeV, however, the widths of these states are 722.8 MeV and 504.7 MeV which are larger than the level spacing suggesting that these states are not discrete and expected to appear as an inseparable structure in any observations.
The all nuclear states shown in Table~\ref{table:BE-W} have the same features.
In contrast, the atomic $1s$ and $2s$ states, which are the 3rd and 4th $s-$states in the solutions of the Klein-Gordon equation, exhibit the level spacing of 174.6 keV with the widths of these states of 57.9 keV and 12.4 keV, respectively.
The widths are smaller enough than the level spacing even for the deepest atomic states and this fact indicates that the two levels are sufficiently separated and can be observed clearly with almost no overlap by appropriate methods.

Then, we consider the whole structure of the energy levels including the different angular momentum $l$ states.
The all results for the nuclear states in Table~\ref{table:BE-W} show the similar level structures of that of the harmonic oscillator potential and the approximate degeneracy between the pairs of $2s$ and $3d$, $3p$ and $4f$, and so on.
For the atomic states, the level structures roughly resemble that of the point charge Coulomb potential and show the approximate degeneracy between the levels with the same principal quantum number with the tendency of the smaller binding energies for the states with the smaller $l$ because of the repulsive effects around the region of the nucleus due to the existence of the nuclear states.
The finite charge density distributions also have the repulsive contribution around the region of the nucleus compared to the point charge Coulomb potential.
One may think that these degeneracies may arise the problems for the separate observations of individual atomic levels.
These states, however, have the different angular momentum quantum numbers and, in principle, can be distinguished by the experiments with the different momentum transfer such as the angular dependence of the energy spectra as in pionic atom formation \cite{Ikeno:2011aa,piAF:2017nha}.
As can be see in Table~\ref{table:BE-W}, all atomic states considered here are found to have small widths and expected to be observed separately by the $({\bar p},p)$ reactions as described below.

We show in Fig.~\ref{fig:12C(pbar,p)11B_pbar_spectra} the calculated formation spectra of ${\bar p}-^{11}$B bound systems by the $^{12}$C$({\bar p},p)$ reactions at $p_{\bar p}=1$ GeV/c as functions of the emitted proton kinetic energies at $\theta_{\rm lab}=0^{\circ}~{\rm and}~5^{\circ}$.
Here, we use the effective number approach with the phenomenological ${\bar p}-$nucleus optical potential of Eqs.~\eqref{eq:Optical} and \eqref{eq:b0}.
The vertical line indicates the threshold energy of the formation of the $ {\bar p}$ bound states in the ground state of $^{11}$B, and the right side of the threshold indicates the region of the antiproton bound states formation and the left side is that of the contributions of the quasi-free antiproton and the nuclear excited states.
We neglect the recoil energy because of the small momentum transfer of this reaction.
We include the contributions of 2 single particle states of proton in $^{12}$C and 14 (4 nuclear and 10 atomic) bound states of antiproton in $^{11}$B as shown in Table~\ref{table:Sp} and \ref{table:BE-W}, and add up all $2\times14=28$ combinations of these states to get total spectrum shown in Fig.~\ref{fig:12C(pbar,p)11B_pbar_spectra}.
The quasi-free antiproton production processes are not evaluated in this article.
In Fig.~\ref{fig:12C(pbar,p)11B_pbar_spectra}(a), the spectrum exhibits three prominent peaks and four minor structures at $\theta_{\rm lab}=0^{\circ}$, which are all attributed to proton knockout from $1p_{3/2}$ state.
The prominent peaks originate from the formation of antiprotonic atoms in $p-$states, and the minor structures from those in $s-$states.
The enhancement of the contributions of the combination of the $p-$states of antiproton and proton is attributed to the small momentum transfer about $q=22$ MeV/c at $\theta_{\rm lab}=0^{\circ}$ in Fig.~\ref{fig:12C(pbar,p)11B_pbar_spectra}(a).
All contributions of proton $s_{1/2}$ state and $ {\bar p}-$nuclear states are included in Fig.~\ref{fig:12C(pbar,p)11B_pbar_spectra} and found to appear only as the tiny and almost flat background of the peaks because of their significantly larger widths.
The contributions of $d$ and $f$ states of $ {\bar p}$ atom can not be seen in the spectrum because of their small formation rate evaluated by the effective numbers.
They could be seen as thin, needle-like peaks only when the extremely high energy resolution, which is comparable to their natural widths shown in Table~\ref{table:BE-W}, is achieved in future experiments.
In Fig.~\ref{fig:12C(pbar,p)11B_pbar_spectra}(b), we can see the spectrum at $\theta_{\rm lab}=5^{\circ}$ where the momentum transfer of the reaction is about $q=89$ MeV/c.
Due to the larger momentum transfer at this angle, the contributions from the antiproton $p-$states become relatively smaller, and those from the $s-$states are more pronounced as can be seen more clearly as peaks.
This behavior of the spectrum can be understood by the matching condition of the momentum transfer and the angular momentum transfer written as,
\begin{equation}
	\Delta L = R \times q,
\end{equation}
where $\Delta L$ is the angular momentum transfer to the target nuclear system in the $( {\bar p},p)$ reaction, and $R$ the root mean square radius of $^{11}$B evaluated as $R=2.42$ fm by the density distribution shown in Table~\ref{table:ADATA}.
For example, the $ {\bar p}$-atomic $p$-states are largely populated at the forward spectrum for smaller momentum transfer because the angular momentum transfer $\Delta L$ can be smaller by coupling to the proton $p$-states than the $ {\bar p}$-atomic $s$-states.
For the angles with the larger momentum transfer, the subcomponents with large $\Delta L$ are more enhanced.
Thus, we can expect to observe the antiprotonic atoms with different angular momentum at finite angles by changing the momentum transfer of the reaction.

\begin{figure}[htbp]
	\centering
	\includegraphics[width=0.7\columnwidth]{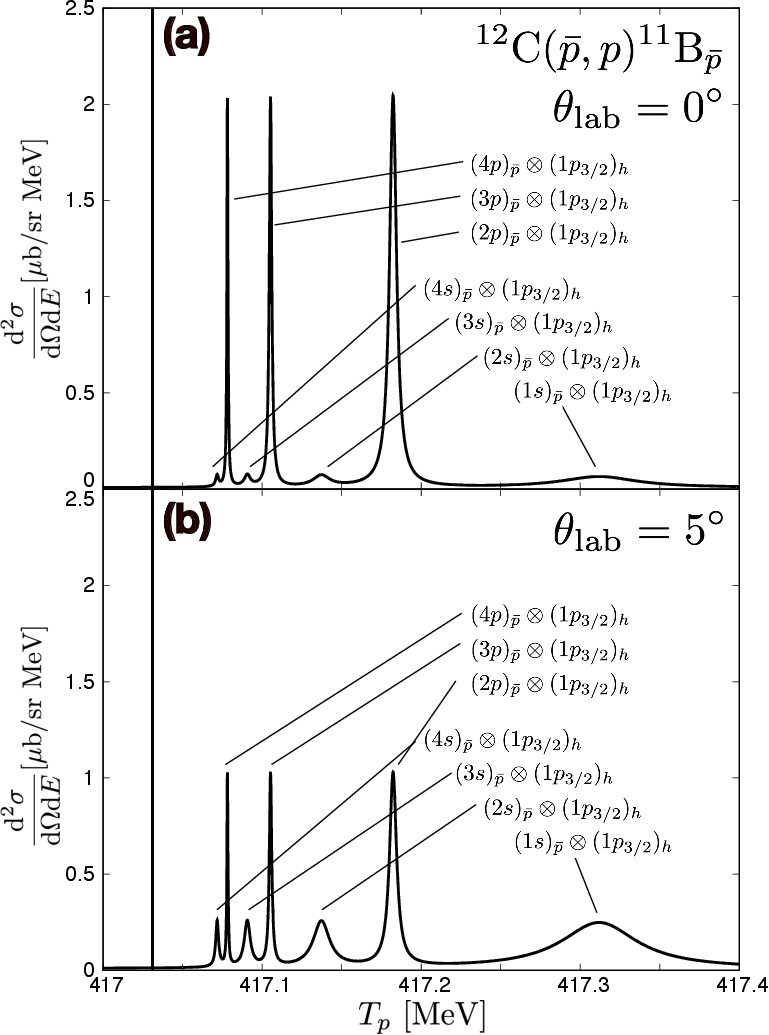}
	\caption{Formation cross sections of $ {\bar p}-^{11}$B bound systems in $^{12}\text{C}(\bar{p},p)$ reactions at $p_{\bar{p}}=1$ GeV/c plotted as functions of the emitted proton kinetic energies at (a) $\theta_{\rm lab}=0^{\circ}$ and (b) $\theta_{\rm lab}=5^{\circ}$.
The vertical solid line indicates the threshold energy of the $ {\bar p}$ bound states formation with the ground state of the daughter nucleus.
Main subcomponents of the spectra are indicated in the figure as the combinations of the $ {\bar p}$ atomic state $(n\ell)_{ {\bar p}}$ and the proton-hole state $(n\ell_{j})_{h}$ in the final system.
The contributions from $ {\bar p}$ nuclear states and $1s_{1/2}$ proton-hole state are included, which do not show any structures in these figures because of their large widths.
The contributions of $d$ and $f$ states of $ {\bar p}$ atom are not seen in the spectrum (see text).}
	\label{fig:12C(pbar,p)11B_pbar_spectra}
\end{figure}

We show in Fig.~\ref{fig:16O(pbar,p)15N_pbar_spectra} the calculated results of the $^{16}{\rm O} ( {\bar p},p)$ reactions for the formation of $ {\bar p}-^{15}$N bound states at (a) $\theta_{\rm lab}=0^{\circ}$ and (b) $\theta_{\rm lab}=5^{\circ}$.
All structures in Fig.~\ref{fig:16O(pbar,p)15N_pbar_spectra} correspond to contributions from the proton knockout from the $1p_{1/2}$ state.
Due to the larger nuclear radius and the smaller atomic Bohr radius of the $ {\bar p}-^{15}$N system compared to those of $^{11}$B, the widths of the $ {\bar p}$ bound states tend to be larger for $^{15}$N.
Even so, the states are still discrete and can be seen as the separated peaks as shown in Fig.~\ref{fig:16O(pbar,p)15N_pbar_spectra}.
As in the case of $^{12}$C target, only atomic $p$ states formation is enhanced at $\theta_{\rm lab}=0^{\circ}$, and the atomic $s$ states become relatively larger at $\theta_{\rm lab}=5^{\circ}$ as expected by the matching condition.
In addition, the contributions from protons in $1p_{3/2}$ state show the almost same peak structures as those from $1p_{1/2}$ shown in Fig.~\ref{fig:16O(pbar,p)15N_pbar_spectra} since the width of the $1p_{3/2}$ proton-hole state is small as shown in Table~\ref{table:Sp}.
The height of the peaks with $1p_{3/2}$ hole are almost twice of those in Fig.~\ref{fig:16O(pbar,p)15N_pbar_spectra} simply because twice as many protons exist in the $1p_{3/2}$ state.
The structures from the $1p_{3/2}$ state are not shown in Fig.~\ref{fig:16O(pbar,p)15N_pbar_spectra} since they appear at 6.3 MeV lower emitted proton energies than the peaks from the $1p_{1/2}$ protons because of the excitation energy of the $^{15}$N.
The contributions from $ {\bar p}-$nuclear states and $1s_{1/2}$ proton-hole state are included in Fig.~\ref{fig:16O(pbar,p)15N_pbar_spectra} and found not to show any structure again because of their large widths.

\begin{figure}[htbp]
\centering
\includegraphics[width=0.7\columnwidth]{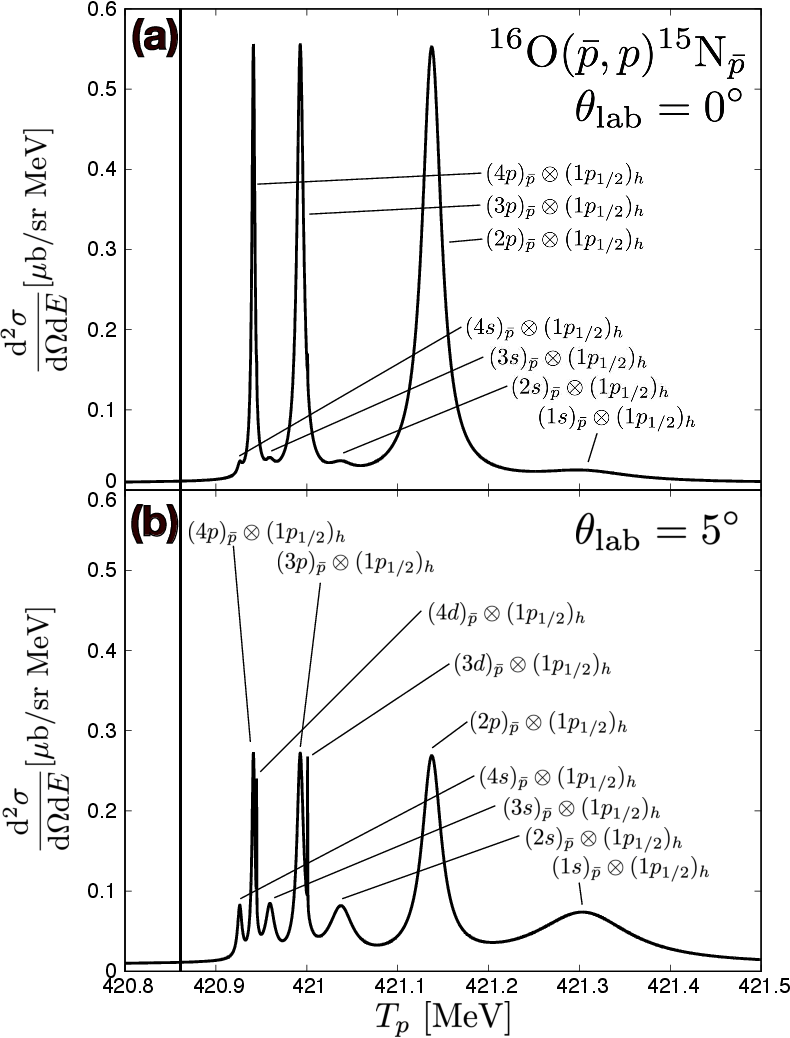}
\caption{Same as Fig.~\ref{fig:12C(pbar,p)11B_pbar_spectra} except for $ {\bar p}-^{15}{\rm N}$ bound systems in $^{16}{\rm O}( {\bar p},p)$ reactions.
The contributions from the $(1p_{3/2})_{h}$ state are not shown in these figures because they appear at 6.3 MeV lower proton energies in the area outside the figure range (see text).}
\label{fig:16O(pbar,p)15N_pbar_spectra}
\end{figure}

Finally, we will consider here the possibility to observe the deepest atomic $1s$ state of $ {\bar p}$ as the dominant peak structure in the $( {\bar p},p)$ spectrum.
To obtain the large formation rate of the bound states, it will be better to consider the reaction with smaller momentum transfer since, in general, reactions with larger momentum transfer tend to yield smaller cross sections than those with smaller momentum transfer.
For $^{12}$C and $^{16}$O target cases, it will be necessary to consider finite scattering angles in order to observe the $1s$ atomic state of the antiproton as shown in  Figs.~\ref{fig:12C(pbar,p)11B_pbar_spectra} and \ref{fig:16O(pbar,p)15N_pbar_spectra} since the valence protons exist in $p$ state in the target.
Thus, we consider the appropriate target nucleus with proton in the valence $s$ state to form the $ {\bar p}$ atom $1s$ state in the reaction with small momentum transfer.
In addition, the daughter nucleus with a proton-hole in the $s$ state must be stable with very small width to observe the peak structure for the $ {\bar p}-$atomic state formation as we have also seen in the $^{12}$C and $^{16}$O target cases.
As the suitable target nucleus for our purpose, we consider $^{31}$P in this article.
The proton pick-up from valence $2s_{1/2}$ state in $^{31}$P provides the stable ground state of $^{30}$Si and, thus, $^{31}$P is considered to be very much suited for the formation of $ {\bar p}$ atom $s$ states in the $^{31}{\rm P}( {\bar p},p)$ reactions at forward angles.

In order to predict a realistic shape and magnitude of the spectrum for the $( {\bar p},p)$ reactions for antiprotonic state formation, we should consider realistic ground state configurations for the target nuclei, the excitation energies and widths of the nuclear states and the relative excitation strengths leading to the excited states of the daughter nuclei in case they are necessary \cite{Ikeno:2013wza}.
The separation energy of $2s_{1/2}$ state of $^{31}$P is 7.3 MeV \cite{chen:2022} and the first excitation energy of $^{30}$Si is 2.2 MeV corresponding to the state of $^{31}$P with $(1d_{5/2})^{-1}$ proton-hole state.
This excitation energy is sufficiently larger than the binding energy of the $1s$ atom of $ {\bar p}$ in $^{30}$Si, which is 1100.8 keV as shown in Table~\ref{table:BE-W}, and the $ {\bar p}$ atom formation spectrum with the $^{30}$Si ground state is expected to be well-separated from the contributions of other excited states of $^{30}$Si in the energy spectrum.
Therefore, in this study, only the valence $2s_{1/2}$ state is considered as the proton single particle state in $^{31}$P.
Then, we consider the occupation probability of the $2s_{1/2}$ state to obtain the realistic magnitude of the  formation cross section.
The occupation probability of the $2s_{1/2}$ state is reported to be 0.40 in Ref.~\cite{Wesseling:1992xra} which was obtained from the analysis of the data of electron-induced proton knock-out $(e,e'p)$ reaction.
The calculated effective numbers for the $ {\bar p}$ atom formation with proton pick-up from the $2s_{1/2}$ state are normalized by this occupation probability.
We show in Fig.~\ref{fig:31P(pbar,p)30Si_pbar_spectra} the formation spectra of $ {\bar p}-^{30}$Si bound systems by the $^{31} {\rm P}( {\bar p},p)$ reaction at $p_{ {\bar p}}=1$ GeV/c.
The contributions from $ {\bar p}$-nuclear states are included in Fig.~\ref{fig:31P(pbar,p)30Si_pbar_spectra} and found not to show any structure, as in previous cases.
We find that we can clearly see the peak structure for the formation of atomic $s$ states in $^{30}$Si, and even the deepest atomic $1s$ state still appears clearly as the prominent discrete peak.

\begin{figure}[htbp]
\centering
\includegraphics[width=0.7\columnwidth]{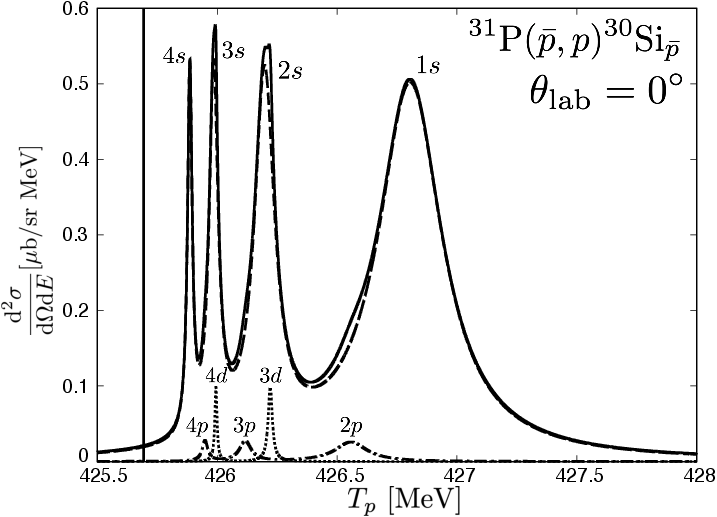}
\caption{Formation cross sections of $ {\bar p}-^{30}$Si (Ground State) bound systems in $^{31}{\rm P}( {\bar p},p)$ reactions at $p_ {\bar p}=1$ GeV/c plotted as functions of the emitted proton kinetic energies at $\theta_{\rm lab}=0^{\circ}$.
Contributions only from the valence $2s_{1/2}$ proton state in $^{31}$P are taken into account.
The contributions from $ {\bar p}$-nuclear states are included, which do not show any structures.
The vertical solid line indicates the threshold energy of $ {\bar p}$ bound state formation.
The quantum numbers of $ {\bar p}$ atomic state $(n\ell)$ are indicated at each peak position in the figure.
Dashed, dash-dotted, and dotted lines show the formation spectra of the $ {\bar p}$ atom of $s$, $p$, and $d$ states, respectively.
The total spectrum is shown by the solid line.
The occupation probability 0.4 for the $2s_{1/2}$ state in $^{31}$P is taken into account.}
\label{fig:31P(pbar,p)30Si_pbar_spectra}
\end{figure}

\section{Summary and Conclusion}
We investigate  in this article the structure and formation of the deeply bound $ {\bar p}$ atoms theoretically.
We use the phenomenological $ {\bar p}$-nucleus optical potential to study the $ {\bar p}$ bound states in nucleus.
We obtain the binding energy levels and wave functions of the $ {\bar p}$ atoms in nucleus, and we confirm that the $ {\bar p}$ atomic states have the larger level spacing than their natural widths and exist as discrete states even for the deepest $1s$ atomic state.
On the other hand, the $ {\bar p}$ nuclear states bounded almost inside the nucleus are found to have very large widths and only exist in the form like continuum state.

We then consider in detail the formation reaction of the deeply bound $ {\bar p}$ atoms which can not be populated by the X-ray observation because of the nuclear absorption effects in the de-excitation processes of the $ {\bar p}$ atom.
We conclude that the $( {\bar p},p)$ reaction is very suited for the formation of the $ {\bar p}$ atoms because of the small momentum transfer at forward angles for all incident energies.
We evaluate the $ {\bar p}$ atom formation cross section of the $( {\bar p},p)$ reaction theoretically for the $^{12}$C, $^{16}$O, and $^{31}$P target nuclei.
We use the effective number approach and find that the atomic states including the deepest $1s$ state can be seen as discrete peak structures in the $( {\bar p},p)$ spectrum.
And we find that the $^{31}$P target is appropriate for the formation of the $ {\bar p}$ atom of $s$ states accompanied by the emission of the proton in the $2s_{1/2}$ state in $^{31}$P.

As for the experimental feasibilities, we first consider the peak structures associated with atomic $p$ states for the $^{12}$C and $^{16}$O targets show in the theoretical spectra.
Because the natural widths of these $p$ states are very small, their experimental separation would require an energy resolution well below 100 keV, making their observation in in-flight antiproton-induced reactions experimentally challenging.
In contrast, for the $^{31}$P target, the corresponding atomic $s$ states have relatively larger widths and the requirements on the energy resolution are less severe, suggesting that the atomic $s$ states may be more accessible in formation reactions.
We also mention here the effects of the backgrounds due to the $ {\bar p}$-nuclear states formation with huge widths.
Since the wavefunctions of these states resemble the proton wavefunctions in nucleus, there may be concern that the effective number $N_{\rm eff}$ of $ {\bar p}$-nuclear states defined in Eq.~\eqref{eq:Neff} could become significantly larger than those of $ {\bar p}$-atomic states.
In reality, the distortion effects are stronger toward the center of the nucleus, so $N_{\rm eff}$ of $ {\bar p}$-nuclear states does not become that significant.
In addition, the widths of these nuclear states are much larger than those of atomic states as shown in Table~\ref{table:BE-W}.
The small momentum transfer of the reaction also helps to enhance the $ {\bar p}$-atomic contributions.
Therefore, the peak height of the nuclear state formation spectra, which roughly scale as $\sim N_{\rm eff}/\Gamma_{ {\bar p}}$, is significantly smaller than that of the atomic state formation in the present model.
Thus, the background  from the $ {\bar p}$-nuclear state is negligible in the calculated spectra in this article.
Further theoretical studies require the application of the Green's function method for more appropriate descriptions of the $ {\bar p}$-nuclear states with large widths \cite{Yamagata:2007cp}.
The experimental efforts for the $ {\bar p}$-nuclear state search were performed by the $( {\bar p},p)$ reaction at a lower energy \cite{Garreta:1984ru,Aslanides:1987vx}, which reported the negative results.
These results will provide the useful information for the future search of the deeply bound $ {\bar p}$ atoms with significantly narrower widths than the $ {\bar p}$-nucleus state.

In conclusions, in our theoretical study we find the existence of the quasi-stable discrete deeply bound $ {\bar p}$ atoms in nuclei, and  we show that the $( {\bar p},p)$ reactions are one of the best method for the formation of the deeply bound $ {\bar p}$ atoms in the laboratory.
We believe that our study largely contributes to the advancement of the research of the $ {\bar p}$ physics at finite density.

\section*{Acknowledgment}
We appreciate the fruitful discussions with Hiroyuki Fujioka.
The work was partly supported by JSPS KAKENHI Grant Number JP22K03607 and JP23K03417.

\let\doi\relax

\end{document}